\documentclass[12pt]{article}

\usepackage{amsmath,amssymb,amscd,amsbsy,amsgen,amsopn,amstext,
amsxtra}

\usepackage[dvips]{graphicx}

\begin{document}

\begin{flushright}
\end{flushright}

\vskip 0.5 truecm

\begin{center}
{\Large{\bf Geometric phases, gauge symmetries and ray 
representation }}
\end{center}
\vskip .5 truecm
\centerline{\bf  Kazuo Fujikawa }
\vskip .4 truecm
\centerline {\it Institute of Quantum Science, College of 
Science and Technology}
\centerline {\it Nihon University, Chiyoda-ku, Tokyo 101-8308, 
Japan}
\vskip 0.5 truecm


\begin{abstract}
The conventional formulation of  the non-adiabatic 
(Aharonov-Anandan) phase is based on the equivalence class 
$\{e^{i\alpha(t)}\psi(t,\vec{x})\}$ which is not a symmetry of the Schr\"{o}dinger equation. This equivalence class when
understood as defining generalized rays in the Hilbert space
is not generally consistent with the superposition principle 
in interference and polarization phenomena. The hidden local
 gauge symmetry, which arises from the arbitrariness of the 
choice of coordinates in the functional space, is then  proposed 
as a basic gauge symmetry in the non-adiabatic phase. This 
re-formulation reproduces all the successful aspects of the 
non-adiabatic phase in a manner manifestly consistent with the 
conventional notion of rays and the superposition 
principle. The hidden local symmetry is thus identified as the 
natural origin of the gauge symmetry in both of the adiabatic 
and non-adiabatic phases in the absence of gauge fields, and it 
allows a unified treatment of  all the geometric phases. The non-adiabatic phase may well be regarded as a special case of 
the adiabatic phase in this re-formulation, contrary to the 
customary understanding of the adiabatic phase as a special
case of the non-adiabatic phase.
Some explicit examples of geometric  phases are discussed to 
illustrate this re-formulation.   
\end{abstract}


\section{Introduction}
The study of geometric phases is an attempt to understand 
quantum mechanics better. The geometric phases have been mainly 
analyzed by using the adiabatic 
approximation~\cite{berry}-\cite{anandan}, though the 
processes slightly away from  adiabaticity have been considered 
in \cite{berry2}. A definition of the non-adiabatic phase, which
 is closely related to the adiabatic phase but without assuming
adiabaticity, has been proposed in~\cite{aharonov, anandan2}. A 
generalization of geometric phases for noncyclic evolutions has 
also been proposed~\cite{samuel}, where the old idea of 
Pancharatnam~\cite{pancharatnam} played an important role. These
earlier works have been further elaborated by various authors,
for example, in Refs.~\cite{review,aitchison, mukunda,  
garcia, mostafazadeh} and references therein.

It has been recently shown~\cite{fujikawa} that gauge symmetries
 involved in the adiabatic (Berry) phase and the non-adiabatic 
(Aharonov-Anandan) phase are quite different by using a second 
quantized formulation~\cite{fujikawa2}. In this formulation the 
hidden local gauge symmetry, which appears as a result of the 
arbitrariness of the phase choice of the complete orthonormal 
basis set in field theory, provides a basis for the
parallel transport and holonomy in the analysis of 
adiabatic phases~\cite{simon}; this local symmetry itself is 
exact regardless of adiabatic or non-adiabatic processes.

In the present paper, we analyze the physical implications of 
these two different gauge symmetries appearing in the 
definitions of geometric phases. The gauge symmetry in the 
non-adiabatic phase is based on the equivalence 
class~\cite{aharonov, anandan2, samuel}
\begin{eqnarray}
\{e^{i\alpha(t)}\psi(t,\vec{x})\}
\end{eqnarray}
instead of constant phases in the conventional definition of 
rays in the Hilbert space~\cite{dirac, streater}. Since the 
Schr\"{o}dinger equation is  not invariant under the equivalence
class (1), one may consider an equivalence class of 
Hamiltonians
$\{\hat{H}-\hbar\partial_{t}\alpha(t)\}$. The gauge symmetry 
means an assignment  of the physical significance to those 
quantities invariant under gauge transformations. A convenient 
way to identify a gauge invariant quantity is to impose  
the parallel transport condition
\begin{eqnarray}
\int d^{3}x \bar{\psi}(t,\vec{x})^{\dagger}i\partial_{t}
\bar{\psi}(t,\vec{x})=0
\end{eqnarray}
by choosing a suitable parameter $\alpha(t)$ in 
$\bar{\psi}(t,\vec{x})=e^{i\alpha(t)}\psi(t,\vec{x})$. This 
$\bar{\psi}(t,\vec{x})$ is written as 
\begin{eqnarray}
\bar{\psi}(t,\vec{x})=\exp[i\int_{0}^{t}dt 
\int d^{3}x \psi(t,\vec{x})^{\dagger}i\partial_{t}
\psi(t,\vec{x})]\psi(t,\vec{x})
\end{eqnarray}
and it is invariant up to a constant phase factor for any 
choice of $\psi(t,\vec{x})$ in the above equivalence class; the 
factor on the exponential plays a role of gauge field.  
This $\bar{\psi}(t,\vec{x})$ thus has the same property as the 
conventional Schr\"{o}dinger amplitude $\psi(t,\vec{x})$ under 
the hidden local 
symmetry~\cite{fujikawa}. However, $\bar{\psi}(t,\vec{x})$ is 
non-local and non-linear in $\psi(t,\vec{x})$ and a linear 
superposition of $\psi(t,\vec{x})$ does not lead to a linear 
superposition of $\bar{\psi}(t,\vec{x})$ in general. 
The variable $\bar{\psi}(t,\vec{x})$ also satisfies  
\begin{eqnarray}
&&i\hbar\frac{\partial}{\partial t}\bar{\psi}(t,\vec{x})
\nonumber\\
&&=[\hat{H}(t)- \int d^{3}x\bar{\psi}^{\dagger}
\hat{H}(t)\bar{\psi}/\int d^{3}x|\bar{\psi}|^{2}]
\bar{\psi}(t,\vec{x})
\end{eqnarray}
if $\psi(t,\vec{x})$  satisfies the ordinary linear 
Schr\"{o}dinger equation. Even in the adiabatic 
limit, a linear superposition of two  independent solutions 
of (4) does not generally satisfy (4).

We examine to what extent the equivalence class 
(1) is regarded  as defining a generalization of
conventional rays, and it is shown that the generalized rays 
thus defined are not generally consistent with the superposition 
principle both in the interference and polarization phenomena. 
It is also explained that the equivalence class (1) in the 
non-adiabatic phase is not reduced to the gauge symmetry in the 
adiabatic phase even in the adiabatic limit. As a result, 
these two gauge symmetries give rise to different constraints in
the measurements of the adiabatic phase by interference.

To reconcile these complications with the attractive idea of the 
non-adiabatic phase, we suggest a 
re-formulation of the non-adiabatic phase on the basis of hidden 
local gauge symmetry arising from the arbitrariness of the 
choice of  coordinates in the functional space~\cite{fujikawa}. 
The hidden local gauge symmetry keeps $\psi(t,\vec{x})$ 
invariant up to a constant phase, namely, $\psi(t,\vec{x})
\rightarrow e^{i\alpha(0)}\psi(t,\vec{x})$ in contrast to 
(1). We show that this 
re-formulation reproduces all the successful aspects of the 
non-adiabatic phase in a way manifestly consistent with the 
conventional notion of rays and the superposition principle.
 The hidden local gauge symmetry  
controls both of the adiabatic and non-adiabatic phases. We 
thus  understand the natural origin of the gauge symmetry in 
geometric phases, which appears even in the absence of gauge 
fields.  
Conceptually, our re-formulation identifies both of the 
adiabatic and non-adiabatic phases as associated with the 
parallel transport and holonomy of an orthonormal basis set, 
rather than the Schr\"{o}dinger amplitude itself, which 
specifies the coordinates of the functional space.

In the present paper, we first recapitulate the basic aspects of
 the hidden local gauge symmetry and the non-adiabatic phase in 
Sections 2 and 3.
The consistency of the equivalence class (1), when understood as 
a generalized notion of rays, with the 
superposition principle is examined in Section 4. 
We then present the re-formulation of the non-adiabatic phase 
on the basis of hidden local symmetry in Section 5 and discuss 
some explicit examples of geometric phases to illustrate the 
re-formulation in Section 6.
  
\section{Hidden local gauge symmetry}

We start with the generic hermitian Hamiltonian 
$\hat{H}=\hat{H}(\hat{\vec{p}},\hat{\vec{x}},X(t))$
for a single particle theory in the  background 
variable $X(t)=(X_{1}(t),X_{2}(t),...)$.
The path integral for this theory for the time interval
$0\leq t\leq T$ in the second quantized 
formulation is given by 
\begin{eqnarray}
Z&=&\int{\cal D}\psi^{\dagger}{\cal D}\psi
\exp\{\frac{i}{\hbar}\int_{0}^{T}dtd^{3}x[
\psi^{\dagger}(t,\vec{x})i\hbar\frac{\partial}{\partial t}
\psi(t,\vec{x})\nonumber\\
&&-\psi^{\dagger}(t,\vec{x})
\hat{H}(\frac{\hbar}{i}\frac{\partial}{\partial\vec{x}},
\vec{x},X(t))\psi(t,\vec{x})] \}.
\end{eqnarray}
We then define a complete set of eigenfunctions
\begin{eqnarray}
&&\hat{H}(\frac{\hbar}{i}\frac{\partial}{\partial\vec{x}},
\vec{x},X(t))v_{n}(\vec{x},X(t))
={\cal E}_{n}(X(t))v_{n}(\vec{x},X(t)), \nonumber\\
&&\int d^{3}x v^{\dagger}_{n}(\vec{x},X(t))v_{m}(\vec{x},X(t))
=\delta_{n,m},
\end{eqnarray}
and expand the classical field $\psi(t,\vec{x})$ in the path integral which is a Grassmann number for a fermion, for example,
as
\begin{eqnarray}
\psi(t,\vec{x})=\sum_{n}b_{n}(t)v_{n}(\vec{x},X(t)).
\end{eqnarray}
We then have
${\cal D}\psi^{\dagger}{\cal D}\psi
=\prod_{n}{\cal D}b_{n}^{\star}
{\cal D}b_{n}$
and the path integral in the second quantized formulation is 
written as 
\begin{eqnarray}
&&Z=\int \prod_{n}{\cal D}b_{n}^{\star}{\cal D}b_{n}
\exp\{\frac{i}{\hbar}\int_{0}^{T}dt[
\sum_{n}b_{n}^{\star}(t)i\hbar\frac{\partial}{\partial t}
b_{n}(t)\nonumber\\
&&+\sum_{n,m}b_{n}^{\star}(t)
\langle n|i\hbar\frac{\partial}{\partial t}|m\rangle
b_{m}(t)
-\sum_{n}b_{n}^{\star}(t){\cal E}_{n}(X(t))b_{n}(t)] \}
\nonumber\\
&&
\end{eqnarray}
where the second term in the action, which is defined by 
\begin{eqnarray}
\int d^{3}x v^{\dagger}_{n}(\vec{x},X(t))
i\hbar\frac{\partial}{\partial t}v_{m}(\vec{x},X(t))
\equiv \langle n|i\hbar\frac{\partial}{\partial t}|m\rangle,
\nonumber
\end{eqnarray}
stands for the term
commonly referred to as Berry's phase\cite{berry} and its 
off-diagonal generalization. 
 We take the time $T$ 
as a period of the  variable $X(t)$ in the analysis
of geometric phases, unless stated otherwise. The adiabatic 
process means that $T$ is much larger than the typical
time scale $\hbar/\Delta{\cal E}_{n}(X(t))$.

Translated into the operator formulation,
we thus obtain the effective Hamiltonian (depending on Bose or 
Fermi statistics)
\begin{eqnarray}
\hat{H}_{eff}(t)&=&\sum_{n}\hat{b}_{n}^{\dagger}(t)
{\cal E}_{n}(X(t))\hat{b}_{n}(t)\nonumber\\
&&-\sum_{n,m}\hat{b}_{n}^{\dagger}(t)
\langle n|i\hbar\frac{\partial}{\partial t}|m\rangle
\hat{b}_{m}(t)
\end{eqnarray}
with $[\hat{b}_{n}(t), \hat{b}^{\dagger}_{m}(t)]_{\mp}
=\delta_{n,m}$.
All the information about geometric phases  is included in 
the effective Hamiltonian and thus purely {\em dynamical}.
See also Berry~\cite{berry2} for a related observation. 
When one defines the Schr\"{o}dinger picture 
$\hat{{\cal H}}_{eff}(t)$ by replacing all $\hat{b}_{n}(t)$ by
$\hat{b}_{n}(0)$ in the above $\hat{H}_{eff}(t)$,
the second quantization formula for the evolution operator 
gives rise to~\cite{fujikawa2}  
\begin{eqnarray}
&&\langle m|T^{\star}\exp\{-\frac{i}{\hbar}\int_{0}^{T}
\hat{{\cal H}}_{eff}(t)
dt\}|n\rangle\nonumber\\ 
&&=
\langle m(T)|T^{\star}\exp\{-\frac{i}{\hbar}\int_{0}^{T}
\hat{H}(\hat{\vec{p}}, \hat{\vec{x}},  
X(t))dt \}|n(0)\rangle 
\end{eqnarray}
where $T^{\star}$ stands for the time ordering operation.
The state vectors in the second quantization  on the left-hand 
side are defined by $
|n\rangle=\hat{b}_{n}^{\dagger}(0)|0\rangle$,
and the state vectors on the right-hand side  stand for the 
first quantized states defined by
$\langle\vec{x}|n(t)\rangle=v_{n}(\vec{x},(X(t))$.
Both-hand sides of the above equality (10) are exact, but the 
difference is that the geometric terms, both of diagonal and 
off-diagonal, are explicit in the second quantized formulation 
on the left-hand side.

The probability amplitude which satisfies Schr\"{o}dinger 
equation with $\psi_{n}(\vec{x},0; X(0))=v_{n}(\vec{x};X(0))$
is given by
\begin{eqnarray}
\psi_{n}(\vec{x},t; X(t))=
\langle 0|\hat{\psi}(t,\vec{x})\hat{b}^{\dagger}_{n}(0)|0\rangle
\end{eqnarray}
since $i\hbar\partial_{t}\hat{\psi}=\hat{H}\hat{\psi}$ in the 
present problem. To be explicit, we have 
\begin{eqnarray}
&&\psi_{n}(\vec{x},t; X(t))\\
&&=\sum_{m} v_{m}(\vec{x};X(t))
\langle m|T^{\star}\exp\{-\frac{i}{\hbar}\int_{0}^{t}
\hat{{\cal H}}_{eff}(t)dt\}|n\rangle\nonumber
\end{eqnarray}
by noting that (10) is given by 
$\langle0| \hat{b}_{m}(t)\hat{b}^{\dagger}_{n}(0)|0\rangle$.
This formula is also derived by noting
\begin{eqnarray}
&&\psi_{n}(\vec{x},t; X(t))\nonumber\\
&&=\langle \vec{x}|T^{\star}\exp\{-\frac{i}{\hbar}\int_{0}^{t}
\hat{H}(\hat{\vec{p}}, \hat{\vec{x}},  
X(t))dt \}|n(0)\rangle\nonumber\\ 
&&=\sum_{m} v_{m}(\vec{x};X(t))\nonumber\\
&&\times
\langle m(t)|T^{\star}\exp\{-\frac{i}{\hbar}\int_{0}^{t}
\hat{H}(\hat{\vec{p}}, \hat{\vec{x}},  
X(t))dt \}|n(0)\rangle
\end{eqnarray}
and the relation (10).
In the adiabatic approximation, where we assume the dominance of
 diagonal elements, we have
\begin{eqnarray}
&&\psi_{n}(\vec{x},t; X(t))\\
&&\simeq v_{n}(\vec{x};X(t))
\exp\{-\frac{i}{\hbar}\int_{0}^{t}[{\cal E}_{n}(X(t))
-\langle n|i\hbar\frac{\partial}{\partial t}|n\rangle]dt\}.
\nonumber
\end{eqnarray}

The path integral formula (8) is based on the expansion (7) and 
the starting second-quantized path integral (5) depends only on 
the field
variable $\psi(t,\vec{x})$, not on  $\{ b_{n}(t)\}$
and $\{v_{n}(\vec{x},X(t))\}$ separately. This fact shows that 
our formulation contains an exact hidden local gauge symmetry
which keeps the field variable $\psi(t,\vec{x})$ invariant 
\begin{eqnarray}
&&v_{n}(\vec{x},X(t))\rightarrow v^{\prime}_{n}(t; \vec{x},X(t))=
e^{i\alpha_{n}(t)}v_{n}(\vec{x},X(t)),\nonumber\\
&&b_{n}(t) \rightarrow b^{\prime}_{n}(t)=
e^{-i\alpha_{n}(t)}b_{n}(t), \ \ \ \ n=1,2,3,...,
\end{eqnarray}
where the gauge parameter $\alpha_{n}(t)$ is a general 
function of $t$. 
This gauge symmetry (or substitution rule) states the fact 
that the choice of coordinates in the functional space is 
arbitrary and this symmetry by itself does not give any conservation law. This symmetry is exact under a 
rather mild condition that the basis set (6) is not singular, 
namely, it is exact not only for the adiabatic case but also for
 the non-adiabatic case. Consequently, physical observables 
should always respect this symmetry. Also, by using this local 
gauge freedom, one can choose the phase convention of the basis 
set $\{v_{n}(t,\vec{x},X(t))\}$ at one's will such that the 
analysis of geometric phases becomes simplest.

Our next observation is that $\psi_{n}(\vec{x},t; X(t))$ 
transforms under the hidden local gauge symmetry (15) as
\begin{eqnarray} 
\psi^{\prime}_{n}(\vec{x},t; X(t))=e^{i\alpha_{n}(0)}
\psi_{n}(\vec{x},t; X(t))
\end{eqnarray}
{\em independently} of the value of $t$. 
This transformation is derived by using the exact 
representation (11), and it implies that 
$\psi_{n}(\vec{x},t; X(t))$ is a physical object since 
$\psi_{n}(\vec{x},t; X(t))$  stays in the same 
ray~\cite{dirac, streater} 
under an arbitrary hidden local gauge transformation.
This transformation is explicitly checked for the adiabatic
approximation (14) also.

The product
$\psi_{n}(\vec{x},0; X(0))^{\dagger}\psi_{n}(\vec{x},T; X(T))$
is thus manifestly independent of the choice of the phase 
convention of the basis set $\{v_{n}(t,\vec{x},X(t))\}$. 
For the adiabatic formula (14), the gauge invariant quantity
 is given by
\begin{eqnarray}
&&\psi_{n}(\vec{x},0; X(0))^{\dagger}\psi_{n}(\vec{x},T; X(T))
\nonumber\\
&&=v_{n}(0,\vec{x}; X(0))^{\dagger}v_{n}(T,\vec{x};X(T))
\nonumber\\
&&\times\exp\{-\frac{i}{\hbar}\int_{0}^{T}[{\cal E}_{n}(X(t))
-\langle n|i\hbar\frac{\partial}{\partial t}|n\rangle]dt\}.
\end{eqnarray}
We then observe that by choosing the hidden gauge such that 
$v_{n}(T,\vec{x};X(T))=v_{n}(0,\vec{x}; X(0))$, the prefactor 
$v_{n}(0,\vec{x}; X(0))^{\dagger}v_{n}(T,\vec{x};X(T))$ becomes 
real and positive. Note that we are assuming the cyclic 
evolution of the external parameter, $X(T)=X(0)$. Then the 
phase factor in (17) defines a physical quantity uniquely. See 
also Refs.~\cite{garcia, mostafazadeh}. After this gauge 
fixing, the phase in (17) is still invariant under residual 
gauge transformations satisfying the periodic boundary condition
$\alpha_{n}(0)=\alpha_{n}(T)$,
in particular, for $\alpha_{n}(X(t))$.   

A change of the coordinates in the functional space more general 
than (15) is possible~\cite{fujikawa}, and we utilize it to 
describe the non-adiabatic phase later.

\section{Non-adiabatic phase}

We recapitulate the basic aspects of  non-adiabatic 
phases defined by Aharonov and Anandan~\cite{aharonov,anandan2} 
and analyzed further by Samuel and Bhandari~\cite{samuel}. See
also review~\cite{review}.

The analysis in Ref.~\cite{aharonov} starts with the wave 
function satisfying 
\begin{eqnarray}
&&\int d^{3}x \psi(t,\vec{x})^{\dagger}\psi(t,\vec{x})=1,\ \ 
\psi(T,\vec{x})=e^{i\phi}\psi(0,\vec{x})
\end{eqnarray}
with a real constant $\phi$. For simplicity we restrict our 
attention to the unitary time-development as in (18).
The condition (18) then implies the existence of a hermitian 
Hamiltonian
\begin{eqnarray}
i\hbar\frac{\partial}{\partial t}\psi(t,\vec{x})=
\hat{H}(\frac{\hbar}{i}\frac{\partial}{\partial\vec{x}},
\vec{x}, X(t))\psi(t,\vec{x})
\end{eqnarray}
but now the variable $X(t)$ need not be slowly varying.
The mathematical basis of the non-adiabatic phase is
the equivalence class, namely, the identification of all the 
state vectors of the form ("projective Hilbert space") 
\begin{eqnarray}
\{e^{i\alpha(t)}\psi(t,\vec{x})\}.
\end{eqnarray}
Note that they project 
$\psi(t,\vec{x})$ for each $t$, which means local in time
unlike the conventional notion of rays which is based on 
{\em constant} $\alpha$~\cite{dirac, streater}. Since the conventional 
Schr\"{o}dinger equation is not invariant under this 
equivalence class, we may consider an equivalence class of 
Hamiltonians
\begin{eqnarray}
\{ \hat{H} -\hbar\frac{\partial}{\partial t}\alpha(t)\}.
\end{eqnarray}

The equivalence class (20) means that we assign physical 
significance to those quantities invariant under the equivalence 
class.  One can choose a suitable representative
state vector 
$\tilde{\psi}(t,\vec{x})=e^{-i\alpha(t)}\psi(t,\vec{x})$
such that 
\begin{eqnarray}
\tilde{\psi}(T,\vec{x})=\tilde{\psi}(0,\vec{x})
\end{eqnarray}
by choosing $\alpha(T)-\alpha(0)=\phi$. This 
$\tilde{\psi}(t,\vec{x})$ is not invariant under (20), but it 
plays an important role in defining physical quantities.

One can also choose a representative state vector 
$\bar{\psi}(t,\vec{x})=e^{i\alpha(t)}\psi(t,\vec{x})$
such that 
\begin{eqnarray}
&&\int d^{3}x \bar{\psi}^{\dagger}(t,\vec{x})i\partial_{t}
\bar{\psi}(t,\vec{x})\nonumber\\
&&=\int d^{3}x \psi^{\dagger}(t,\vec{x})
i\partial_{t}\psi(t,\vec{x})-\partial_{t}\alpha(t)=0
\end{eqnarray}
namely~\cite{aitchison, mukunda}
\begin{eqnarray}
\bar{\psi}(t,\vec{x})=\exp[i\int_{0}^{t}dt \int d^{3}x
\psi(t,\vec{x})^{\dagger}
i\frac{\partial}{\partial t}\psi(t,\vec{x}) ]
\psi(t,\vec{x})
\end{eqnarray}
up to a constant phase factor $e^{i\alpha(0)}$. The exponential
factor in (24) plays a role of gauge field, and 
under the equivalence class (or gauge transformation)
$\psi(t,\vec{x}) \rightarrow e^{i\alpha(t)}\psi(t,\vec{x})$
one has 
\begin{eqnarray}
\bar{\psi}(t,\vec{x}) \rightarrow e^{i\alpha(0)}
\bar{\psi}(t,\vec{x}).
\end{eqnarray}
This property (25), which is valid independently of the precise 
form of the Hamiltonian in (19) since we use only the property 
(18), implies that $\bar{\psi}$ is a physical gauge invariant 
object up to a constant phase.

The manifestly gauge invariant quantity is then defined by
\begin{eqnarray}
&&\bar{\psi}(0,\vec{x})^{\dagger}\bar{\psi}(T,\vec{x})\\
&&=\psi(0,\vec{x})^{\dagger}\exp[i\int_{0}^{T}dt \int d^{3}x
\psi(t,\vec{x})^{\dagger}
i\frac{\partial}{\partial t}\psi(t,\vec{x}) ]
\psi(T,\vec{x})\nonumber
\end{eqnarray}
by following the prescription (17).
By a suitable gauge transformation 
$\psi(t,\vec{x})\rightarrow \tilde{\psi}(t,\vec{x})=
e^{-i\alpha(t)}\psi(t,\vec{x})$,
we have $\tilde{\psi}(0,\vec{x})=\tilde{\psi}(T,\vec{x})$ as in (22).
The above gauge invariant quantity is then written as
\begin{eqnarray} 
\bar{\psi}(0,\vec{x})^{\dagger}\bar{\psi}(T,\vec{x})
&=&|\tilde{\psi}(0,\vec{x})|^{2}\exp[i\beta]\nonumber\\
&=&|\psi(0,\vec{x})|^{2}
\exp[i\beta]
\end{eqnarray}
with 
\begin{eqnarray}
\beta=\oint dt \int d^{3}x
\tilde{\psi}(t,\vec{x})^{\dagger}
i\frac{\partial}{\partial t}\tilde{\psi}(t,\vec{x})
\end{eqnarray}
which extracts all the information about the phase from the 
gauge invariant quantity. This quantity $\beta$, which is still
invariant under the residual gauge symmetry $\alpha(t)$ with
$\alpha(0)=\alpha(T)$, is called 
``non-adiabatic phase''~\cite{aharonov}. 
  
The Schr\"{o}dinger equation for $\psi(t,\vec{x})=
e^{i\gamma(t)}\tilde{\psi}(t,\vec{x})$ 
\begin{eqnarray}
i\hbar\partial_{t}\psi(t,\vec{x})=\hat{H}\psi(t,\vec{x})
\end{eqnarray}
with
 $\gamma(T)-\gamma(0)=\phi$ implies
\begin{eqnarray}
\hbar\phi&=&\hbar\oint dt \int d^{3}x
\tilde{\psi}(t,\vec{x})^{\dagger}
i\frac{\partial}{\partial t}\tilde{\psi}(t,\vec{x})\nonumber\\
&&-\int_{0}^{T}dt\int d^{3}x
\psi^{\dagger}(t,\vec{x})\hat{H}\psi(t,\vec{x})\nonumber\\
&=&\hbar\beta-\int_{0}^{T}dt\int d^{3}x
\psi^{\dagger}(t,\vec{x})\hat{H}\psi(t,\vec{x}).
\end{eqnarray}
The last term  $\int_{0}^{T}dt\int d^{3}x
\psi^{\dagger}(t,\vec{x})\hat{H}\psi(t,\vec{x})$ on the 
right-hand side is called in~\cite{aharonov} as a ``dynamical 
phase'', though the total phase
$\hbar\phi$ is in fact generated by the Hamiltonian $\hat{H}$
and thus dynamical. Eq.(30) defines  the non-adiabatic
phase and the ``dynamical phase'' simultaneously.

\section{Ray representation and superposition principle}

We examine the physical implications of the 
two different gauge symmetries, the hidden local gauge symmetry 
(15) and the equivalence class (20). The basic correspondence is 
\begin{eqnarray}
v_{n}(\vec{x};X(t)) \leftrightarrow \psi(t,\vec{x})
\end{eqnarray}
with the equivalence classes
\begin{eqnarray}
\{ e^{i\alpha_{n}(t)}v_{n}(\vec{x};X(t))\} \leftrightarrow
\{ e^{i\alpha(t)}\psi(t,\vec{x})\} .
\end{eqnarray}
The physical gauge invariant phases in the cyclic evolution are 
then given by, respectively, (17) and (27).
The two formulations are thus very similar to each other, but 
there is a crucial difference: 
The true correspondence should be
\begin{eqnarray} 
\psi_{n}(\vec{x},t; X(t)) \leftrightarrow \psi(t,\vec{x}), 
\end{eqnarray}
since both of $\psi_{n}(\vec{x},t; X(t))$
in (11) and 
 $\psi(t,\vec{x})$  stand for the Schr\"{o}dinger probability 
amplitudes.  
Note that the probability amplitude need not be a linear 
superposition of basis vectors as is seen in the exact 
expression before approximation in (12). The hidden local 
symmetry (15) gives rise to the conventional 
notion of rays with constant phases, as is seen in (16).  

We would like to understand the physical and conceptual basis 
for postulating
the equivalence class (20). One may understand that the 
equivalence 
class is based on a generalization of the notion of rays in the 
Hilbert space.
We examine this possibility. 
An important property of the Schr\"{o}dinger amplitude is 
that one can consider a superposition of two probability
amplitudes such as 
\begin{eqnarray}
\psi(t,\vec{x})=c_{1}e^{i\alpha_{1}}\psi_{1}(t,\vec{x})
+c_{2}e^{i\alpha_{2}}\psi_{2}(t,\vec{x})
\end{eqnarray}
with two real constants $\alpha_{1}$and $\alpha_{2}$ for the 
solutions of the Schr\"{o}dinger equation
\begin{eqnarray}
i\hbar\partial_{t}\psi_{k}(t,\vec{x})=\hat{H}\psi_{k}(t,\vec{x}),
\ \   k=1,2.
\end{eqnarray}
The superposition  satisfies the same Schr\"{o}dinger equation
$i\hbar\partial_{t}\psi(t,\vec{x})=\hat{H}\psi(t,\vec{x})$
and thus gives a probability amplitude.
This superposition principle is based on the conventional notion
of rays with constant phases.

On the other hand, for the generalized equivalence class
we have 
\begin{eqnarray}
\psi^{\prime}(t,\vec{x})=c_{1}e^{i\alpha_{1}(t)}
\psi_{1}(t,\vec{x})+c_{2}e^{i\alpha_{2}(t)}\psi_{2}(t,\vec{x})
\end{eqnarray}
for the solutions of the Schr\"{o}dinger equation
\begin{eqnarray}
i\hbar\partial_{t}(e^{i\alpha_{k}(t)}\psi_{k}(t,\vec{x}))
=(\hat{H}-\hbar\partial_{t}\alpha_{k}(t))(e^{i\alpha_{k}(t)}\psi_{k}(t,\vec{x})),
\ \   k=1,2\nonumber\\
\end{eqnarray}
The superposition (of linearly independent $\psi_{1}$ and 
$\psi_{2}$) does not satisfy the Schr\"{o}dinger equation
of the general form 
\begin{eqnarray}
i\hbar\partial_{t}\psi^{\prime}(t,\vec{x})=(\hat{H}-
\hbar\partial_{t}\alpha(t))\psi^{\prime}_{k}(t,\vec{x})
\end{eqnarray}
except for the case 
\begin{eqnarray}
\partial_{t}\alpha_{1}(t)=\partial_{t}\alpha_{2}(t).
\end{eqnarray}
If one imposes this condition on the parameters $\alpha(t)$
for any combination of state vectors, the 
generalized ray is reduced to the conventional ray with a new 
Hamiltonian
\begin{eqnarray}
\hat{H}^{\prime}=\hat{H}-\hbar\partial_{t}\alpha_{1}(t).
\end{eqnarray}

Another important consequence of the  equivalence class
of states (20) is that 
one can always choose a representative 
$\bar{\psi}(t,\vec{x})=e^{i\alpha(t)}\psi(t,\vec{x})$
which satisfies the parallel transport condition (23).
Namely,
\begin{eqnarray}
\bar{\psi}(t,\vec{x})=e^{i\int_{0}^{t}dt d^{3}x
\psi(t,\vec{x})^{\dagger}i\partial_{t}\psi(t,\vec{x})}
\psi(t,\vec{x})
\end{eqnarray}
up to a constant phase $e^{i\alpha(0)}$. 
Given the Schr\"{o}dinger equation 
\begin{eqnarray}
i\hbar\partial_{t}\psi(t,\vec{x})=H\psi(t,\vec{x}),
\end{eqnarray}
one has
\begin{eqnarray}
i\hbar\partial_{t}\bar{\psi}(t,\vec{x})&=&(H-
\hbar\partial_{t}\alpha(t))\bar{\psi}(t,\vec{x})\nonumber\\
&=&[H-\int d^{3}x\psi(t,\vec{x})^{\dagger}i\hbar\partial_{t}
\psi(t,\vec{x})]\bar{\psi}(t,\vec{x})\nonumber\\
&=&[H-\int^{}_{}d^{3}x\psi(t,\vec{x})^{\dagger}H\psi(t,\vec{x})]
\bar{\psi}(t,\vec{x})\nonumber\\
&=&[H-\int d^{3}x\bar{\psi}(t,\vec{x})^{\dagger}H
\bar{\psi}(t,\vec{x})]\bar{\psi}(t,\vec{x}).
\end{eqnarray}
Namely, the representative which satisfies the parallel 
transport and gauge invariant conditions satisfies the 
non-linear Schr\"{o}dinger equation\cite{birula, weinberg}. One 
may also write this equation in the form of (4),
which exhibits the symmetry under $\bar{\psi}(t,\vec{x})
\rightarrow Z\bar{\psi}(t,\vec{x})$ with a complex constant 
$Z$~\cite{weinberg}. 

A linear superposition of two representatives 
\begin{eqnarray}
c_{1}\bar{\psi}_{1}(t,\vec{x})
+c_{2}\bar{\psi}_{2}(t,\vec{x})
\end{eqnarray}
of the two equivalence sets of states 
$\{e^{i\alpha_{1}(t)}\psi_{1}(t,\vec{x})\}$ and 
$\{e^{i\alpha_{2}(t)}\psi_{2}(t,\vec{x})\}$, where 
$\psi_{1}(t,\vec{x})$ and $\psi_{2}(t,\vec{x})$ are linearly independent, satisfies the same
(non-linear) Schr\"{o}dinger equation only for 
\begin{eqnarray}
&&\int d^{3}x\psi_{1}(t,\vec{x})^{\star}i\hbar\partial_{t}
\psi_{1}(t,\vec{x})\nonumber\\
&&=\int d^{3}x\psi_{2}(t,\vec{x})^{\star}i\hbar\partial_{t}
\psi_{2}(t,\vec{x})
\end{eqnarray}
which is consistent with $\partial_{t}\alpha_{1}(t)
=\partial_{t}\alpha_{2}(t)$ in (39). 
The superposition of two probability amplitudes which satisfy 
the parallel transport and gauge invariance conditions is 
regarded as the 
Schr\"{o}dinger probability amplitude only under this condition.

The polarization measurement cannot distinguish 
$\psi^{\prime}(t,\vec{x})=e^{i\alpha(t)}\psi(t,\vec{x})$ and 
$\psi(t,\vec{x})$ in the sense that
\begin{eqnarray}
\psi^{\prime}(t,\vec{x})^{\dagger}\vec{\sigma}
\psi^{\prime}(t,\vec{x})=
\psi(t,\vec{x})\vec{\sigma}\psi(t,\vec{x})
\end{eqnarray}
and thus one may regard the generalized rays and the conventional
 rays are equivalent in the analysis of polarization phenomena. 
The situation is however more involved: 
An analysis of the movement of the  polarization vector in the 
constant magnetic field $\vec{B}$ described by, for example, 
\begin{eqnarray}
\hat{H}=-\mu\hbar\vec{\sigma}\vec{B}
\end{eqnarray}
is based on the superposition of two states 
\begin{eqnarray}
&&\psi(t)=\cos\frac{\theta}{2}\psi_{+}(t)+\sin\frac{\theta}{2}
\psi_{-}(t),
\nonumber\\
&&i\hbar\partial_{t}\psi(t)=\hat{H}\psi(t)
\end{eqnarray}
with
\begin{eqnarray}
&& i\hbar\partial_{t}\psi_{\pm}(t)=H\psi_{\pm}(t),\nonumber\\
&&H\psi_{\pm}(t)=\mp\mu\hbar B\psi_{\pm}(t)
\end{eqnarray}
If one uses different representatives in the conventional 
definition of rays with constant phases, 
$\{e^{i\alpha_{1}}\psi_{+}(t) \} \ {\rm and}  
\ \{e^{i\alpha_{2}}\psi_{-}(t) \}$, in (48) the phase factors 
are simply absorbed in the different choice of the superposition 
coefficients $\cos\frac{\theta}{2}$ and $\sin\frac{\theta}{2}$.
 
If one considers the equivalence classes in the notion of
generalized rays
\begin{eqnarray}
\{e^{i\alpha_{1}(t)}\psi_{+}(t) \}, \ \ \ \ 
\{e^{i\alpha_{2}(t)}\psi_{-}(t) \},
\end{eqnarray}
a linear superposition of two representatives 
\begin{eqnarray}
\psi^{\prime}(t)=\cos\frac{\theta}{2} e^{i\alpha_{1}(t)}
\psi_{+}(t)
+\sin\frac{\theta}{2} e^{i\alpha_{2}(t)}\psi_{-}(t)
\end{eqnarray}
does not satisfy the Schr\"{o}dinger equation in general except
for $\partial_{t}\alpha_{1}(t)=\partial_{t}\alpha_{2}(t)$, for 
which the generalized
 ray is reduced to the conventional ray for a modified 
Hamiltonian
$\hat{H}^{\prime}=\hat{H}-\hbar\partial_{t}\alpha_{1}(t)$. 
Incidentally, in 
the present case one cannot maintain the parallel transport 
condition (23) by choosing suitable $\alpha_{1}(t)$ and 
$\alpha_{2}(t)$ in $\{e^{i\alpha_{1}(t)}\psi_{+}(t) \} 
\ {\rm and}\  
\{e^{i\alpha_{2}(t)}\psi_{-}(t) \}$, since for such a case one has to satisfy
$\partial_{t}\alpha_{1}(t)=\partial_{t}\alpha_{2}(t)$ and 
\begin{eqnarray}
&&\hbar\partial_{t}\alpha_{1}(t)
=\int d^{3}x\psi_{+}(t,\vec{x})^{\star}
i\hbar\partial_{t}\psi_{+}(t,\vec{x})=-\mu\hbar B,\nonumber\\
&&\hbar\partial_{t}\alpha_{2}(t)
=\int d^{3}x\psi_{-}(t,\vec{x})^{\star}i\hbar\partial_{t}
\psi_{+}(t,\vec{x})=\mu\hbar B.
\end{eqnarray}
A solution of the Schr\"{o}dinger equation is often written 
as a superposition of two or more other solutions of the 
Schr\"{o}dinger equation. The notion of rays should be 
consistent with such a general situation. Only in the 
conventional definition of rays~\cite{dirac, streater}, one can 
maintain consistency 
and describe the movement of the polarization vector 
consistently.

In the case of the explicit construction of the 
Schr\"{o}dinger amplitude $\psi_{n}(t,\vec{x})$ in (12), one can write
\begin{eqnarray}
&&\psi_{n}(t,\vec{x})=\sum_{m}v_{m}(t,\vec{x})G_{mn}(t),
\nonumber\\
&&i\hbar\partial_{t}\psi_{n}(t,\vec{x})=H\psi_{n}(t,\vec{x}),
\end{eqnarray}
where $G_{mn}(t)$ stands for the unitary evolution operator 
(10) and $\psi_{n}(t,\vec{x})$ is invariant under the hidden local symmetry up to
a constant phase. On the other hand, the quantity 
$\bar{\psi}_{n}(t,\vec{x})$, which is 
invariant under the equivalence class (20) up to a constant 
phase, satisfies
\begin{eqnarray}
&&\bar{\psi}_{n}(t,\vec{x})=\exp[i\int^{t}_{0}\int d^{3}x
\psi_{n}^{\dagger}(t,\vec{x})
i\partial_{t}\psi_{n}(t,\vec{x})]\psi_{n}(t,\vec{x}),\nonumber\\
&&i\hbar\frac{\partial}{\partial t}\bar{\psi}_{n}(t,\vec{x})
\nonumber\\
&&=[\hat{H}(t)- \int d^{3}x\bar{\psi}_{n}^{\star}
\hat{H}(t)\bar{\psi}_{n}]
\tilde{\psi}_{n}(t,\vec{x}),\nonumber\\
&&\int d^{3}x{\bar{\psi}_{n}(t,\vec{x})}^{\dagger}i\partial_{t}
\bar{\psi}_{n}(t,\vec{x})=0.
\end{eqnarray}
From these expressions, one can clearly see the difference 
between the two gauge symmetries. One can also see that the 
gauge symmetry in the non-adiabatic phase is not reduced to the 
hidden local symmetry even in the adiabatic limit. The adiabatic
formula
\begin{eqnarray}
&&\psi_{n}(\vec{x},t; X(t))\\
&&\simeq v_{n}(\vec{x};X(t))
\exp\{-\frac{i}{\hbar}\int_{0}^{t}[{\cal E}_{n}(X(t))
-\langle n|i\hbar\frac{\partial}{\partial t}|n\rangle]dt\}.
\nonumber
\end{eqnarray}
is invariant under the hidden local symmetry (15) up to a 
constant phase, but this symmetry has nothing to do with the 
equivalence
class $\{e^{i\alpha(t)}\psi_{n}(\vec{x},t; X(t))\}$. 

Physically, the basic difference between the two gauge symmetries is that the quantity $\psi_{n}^{\dagger}(0,\vec{x})\psi_{n}(t,\vec{x})$ in (17) invariant under the hidden 
gauge symmetry is directly measurable as the interference term in
\begin{eqnarray}
&&|\psi_{n}^{\dagger}(0,\vec{x})+\psi_{n}(t,\vec{x})|^{2}
\nonumber\\
&&=|\psi_{n}(0,\vec{x})|^{2}
+|\psi_{n}(t,\vec{x})|^{2}+2Re \psi_{n}^{\dagger}(0,\vec{x})
\psi_{n}(t,\vec{x})
\end{eqnarray}
by superposing two beams; for one of the beams one may choose
$\hat{H}=0$  and for the other one may choose $\hat{H}\neq 0$
with the identical kinematical phases which depend on the length 
of the two arms. On the other hand,
the quantity ${\bar{\psi}(0,\vec{x})}^{\dagger}
\bar{\psi}(t,\vec{x})$ in (26) invariant under the equivalence class is not directly measured as the interference term in
\begin{eqnarray}
&&|{\bar{\psi}(0,\vec{x})}^{\dagger}
+\bar{\psi}(t,\vec{x})|^{2}
\nonumber\\
&=&|{\bar{\psi}(0,\vec{x})}^{\dagger}|^{2}
+|\bar{\psi}(t,\vec{x})|^{2}
+2Re {\bar{\psi}(0,\vec{x})}^{\dagger}
\bar{\psi}(t,\vec{x})
\nonumber\\
&=&|{\psi(0,\vec{x})}|^{2}+|\psi(t,\vec{x})|^{2}
\nonumber\\
&&+2Re \{{\psi(0,\vec{x})}^{\dagger}\exp[i\int^{t}_{0}dt
\int d^{3}x
\psi^{\dagger}(t,\vec{x})i\partial_{t}\psi(t,\vec{x})]
\nonumber\\
&&\hspace*{1cm}\times\psi(t,\vec{x})\}
\end{eqnarray}
except for the case
\begin{eqnarray}
\int d^{3}x
\psi^{\dagger}(t,\vec{x})i\partial_{t}\psi(t,\vec{x})
=0
\end{eqnarray}
for {\em all} $t$, which ensures
\begin{eqnarray}
&&i\hbar\partial_{t}\{[\exp[i\int^{t}_{0}dt\int d^{3}x
\psi^{\dagger}(t,\vec{x})i\partial_{t}\psi(t,\vec{x})]
\psi(t,\vec{x})\}\nonumber\\
&&=\hat{H}\{\exp[i\int^{t}_{0}dt\int d^{3}x
\psi^{\dagger}(t,\vec{x})i\partial_{t}\psi(t,\vec{x})]\psi(t)\}.
\end{eqnarray}
Under the condition (58), the interference pattern in
(57) agrees with the pattern in (56) dictated by quantum 
mechanics.
This property (57) of the equivalence class  differs from the 
conventional notion of gauge symmetry where only the gauge invariant quantity is directly measurable.

The last property (57) is also important in the analysis of non-adiabatic 
phases for non-cyclic processes~\cite{samuel} in the 
manner of Pancharatnam, where the measurement of interference 
provides a basic means to define the relative phase. To be 
precise, one can define a unique relative phase in  the 
interference term only
for the {\em integrated} quantity~\cite{fujikawa} in the case 
of the non-cyclic process
\begin{eqnarray}
&&\int d^{3}x{\bar{\psi}(0,\vec{x})}^{\dagger}
\bar{\psi}(t,\vec{x})\nonumber\\
&&=\int d^{3}x{\psi(0,\vec{x})}^{\dagger}\exp[i\int_{0}^{t}
dtd^{3}x\psi(t,\vec{x})^{\dagger}i\partial_{t}
\psi(t,\vec{x})]\psi(t,\vec{x}),
\end{eqnarray}
 but still such a phase is not directly measured by interference.

To summarize the analysis in this section, the 
basis of the equivalence class in the non-adiabatic
 phase may be understood as follows: Given any $\psi(t)$, 
one can consider the equivalence class
\begin{eqnarray}
\{e^{i\alpha(t)}\psi(t)\}
\end{eqnarray}
for the {\em specific} $\psi(t)$, then the notion of the 
equivalence class provides a convenient means to extract the 
geometric property of the very specific $\psi(t)$.  However, the
 equivalence class thus defined has no direct connection to a 
generalization of rays in the Hilbert space, and the physical 
origin of the equivalence class is not clear. Also, the gauge 
invariance is not a criterion of observables, as is 
exemplified by the gauge non-invariance of the Hamiltonian in
(21). In this connection, we mention the
``gauge independent formulation'' on the basis of the density 
matrix~\cite{gauge-ind}. A density matrix for a pure state 
$\psi(t)$
\begin{eqnarray}
\rho(t) =|\psi(t)\rangle\langle\psi(t)|
\end{eqnarray}
is trivially invariant under the equivalence class (20). But 
the density matrix for the pure state does not tell how the 
pure state is formed, and the notion of rays and  the 
superposition principle are crucial in 
the construction of the pure state. Also, the trivial invariance 
of the density matrix under the equivalence class means that 
the equivalence class by itself does not provide any useful 
information for the density matrix. 

\section{ Non-adiabatic phase and hidden local symmetry}

To reconcile the attractive idea of the non-adiabatic phase 
with the conventional notion of rays, we suggest to utilize a 
general unitary transformation of  coordinates
in the functional space~\cite{fujikawa}.
Our observation is very simple: We start with the basic 
assumptions in (18) and (19)
\begin{eqnarray} 
&&i\hbar\partial_{t}\psi(t,\vec{x})=\hat{H}(t)\psi(t,\vec{x}),
\nonumber\\
&&\int d^{3}x \psi^{\dagger}(t,\vec{x})\psi(t,\vec{x})=1,
\nonumber\\
&&\psi(t,\vec{x})=e^{i\phi(t)}\tilde{\psi}(t,\vec{x}),\nonumber\\
&&\tilde{\psi}(T,\vec{x})=\tilde{\psi}(0,\vec{x}),\nonumber\\
&&\phi(T)=\phi, \ \ \ \ \phi(0)=0.
\end{eqnarray}
These assumptions combined with a constraint analogous to (30) 
gives
\begin{eqnarray}
\psi(t,\vec{x})
&=&\tilde{\psi}(t,\vec{x})\exp\{-\frac{i}{\hbar}
[\int_{0}^{t}dt\int d^{3}x \tilde{\psi}^{\dagger}(t,\vec{x})
\hat{H}\tilde{\psi}(t,\vec{x})\nonumber\\
&&\hspace{1.5 cm}-\int_{0}^{t}dt\int d^{3}x 
\tilde{\psi}^{\dagger}(t,\vec{x})i\hbar\partial_{t}\tilde{\psi}(t,\vec{x})]\}
\end{eqnarray} 
which is transformed as $\psi(t,\vec{x})\rightarrow 
e^{i\alpha(t)}\psi(t,\vec{x})$ under the equivalence class of Hamiltonians (21) with fixed $\tilde{\psi}(t,\vec{x})$. Our 
suggestion is rather to regard $\tilde{\psi}(t,\vec{x})$ as 
one of the basis vectors and incorporate the hidden local gauge 
symmetry (15) with fixed $\hat{H}$. Then $\psi(t,\vec{x})$ is 
invariant up to a constant phase under the hidden local 
symmetry, and the hidden local symmetry uniquely 
fixes the non-adiabatic phase as in the case of the adiabatic 
phase in (17).

We now explain the above construction.
We first define a unitary transformation $U(t)$
\begin{eqnarray}
&&w_{n}(t,\vec{x})=\sum_{m}U(t)_{nm}v_{m}(t,\vec{x}),\nonumber\\
&&w_{n}(T,\vec{x})=w_{n}(0,\vec{x}),\nonumber\\
&&\hat{H}(t)v_{m}(t,\vec{x})={\cal E}_{m}(t)v_{m}(t,\vec{x}),
\nonumber\\ 
&&\int d^{3}x v^{\dagger}_{m}(t,\vec{x})v_{n}(t,\vec{x})
=\delta_{m,n}
\end{eqnarray}
by taking the basis set $\{v_{m}(t,\vec{x}) \}$ as a basic 
building block, for the sake of definiteness. We choose the 
unitary transformation such that the first element of the new 
complete orthonormal set $\{w_{n}(t,\vec{x}) \}$ satisfies 
\begin{eqnarray}
w_{1}(t,\vec{x})=\tilde{\psi}(t,\vec{x}),
\end{eqnarray}
which is possible since $\{v_{m}(t,\vec{x}) \}$ forms a complete
orthonormal set. The expansion of the field 
variable in second quantization is then given  by
\begin{eqnarray}
\hat{\psi}(t,\vec{x})&=&\sum_{m}\hat{c}_{m}(t)w_{m}(t,\vec{x})
\nonumber\\
&=&\sum_{m}\hat{b}_{m}(t)v_{m}(t,\vec{x})
\end{eqnarray}
with 
\begin{eqnarray}
\hat{c}_{m}(t)=\sum_{n}\hat{b}_{n}(t)U(t)^{\dagger}_{nm}.
\end{eqnarray}
The variable $\hat{\psi}(t,\vec{x})$ in (67) contains an
exact hidden local symmetry
\begin{eqnarray}
&&w_{m}(t,\vec{x})\rightarrow e^{i\alpha_{m}(t)}w_{m}(t,\vec{x}),
\nonumber\\
&&\hat{c}_{m}(t)\rightarrow e^{-i\alpha_{m}(t)}\hat{c}_{m}(t)  
\end{eqnarray}
with  general functions $\{\alpha_{m}(t)\}$. Following (13),
we define 
\begin{eqnarray}
&&\psi_{n}(t, \vec{x})\nonumber\\
&&=\langle \vec{x}|T^{\star}\exp\{-\frac{i}{\hbar}\int_{0}^{t}
\hat{H}(\hat{\vec{p}}, \hat{\vec{x}},  
X(t))dt \}|n(0)\rangle\nonumber\\ 
&&=\sum_{m} w_{m}(t,\vec{x})\nonumber\\
&&\times
\langle m(t)|T^{\star}\exp\{-\frac{i}{\hbar}\int_{0}^{t}
\hat{H}(\hat{\vec{p}}, \hat{\vec{x}},  
X(t))dt \}|n(0)\rangle\nonumber\\
&&=\sum_{m} w_{m}(t,\vec{x})\nonumber\\
&&\times
\langle m|T^{\star}\exp\{-\frac{i}{\hbar}\int_{0}^{t}
\hat{\cal H}_{eff}(t)dt \}|n\rangle
\end{eqnarray}
where $\hat{\cal H}_{eff}(t)$ in the Schr\"{o}dinger picture
is obtained from 
\begin{eqnarray}
\hat{H}_{eff}(t)&=&\sum_{n,m}\hat{c}_{n}^{\dagger}(t)[
\int d^{3}x w^{\dagger}_{n}(t,\vec{x})\hat{H}(t)w_{m}(t,\vec{x})
\nonumber\\
&&-
\int d^{3}x w^{\dagger}_{n}(t,\vec{x})i\hbar\partial_{t}
w_{m}(t,\vec{x})]\hat{c}_{m}(t)
\end{eqnarray}
by replacing all $\hat{c}_{n}(t)$ by $\hat{c}_{n}(0)$. The state
in the first quantization is defined by 
$\langle\vec{x}|n(t)\rangle=w_{n}(t,\vec{x})$ and the state in 
the second quantization is defined by 
$|n\rangle=c^{\dagger}_{n}(0)|0\rangle$ in (70).

The amplitudes thus defined satisfy 
\begin{eqnarray}
&&i\hbar\partial_{t}\psi_{n}(t,\vec{x})=
\hat{H}(t)\psi_{n}(t,\vec{x}),\nonumber\\
&&\psi_{n}(0,\vec{x})=w_{n}(0,\vec{x}).
\end{eqnarray}
In particular, the amplitude $\psi_{1}(t,\vec{x})$
satisfies
\begin{eqnarray}
&&i\hbar\partial_{t}\psi_{1}(t,\vec{x})=
\hat{H}(t)\psi_{1}(t,\vec{x}),\nonumber\\
&&\psi_{1}(0,\vec{x})=w_{1}(0,\vec{x})=\psi(0,\vec{x}).
\end{eqnarray}
We thus have
\begin{eqnarray}
\psi(t,\vec{x})&=&\psi_{1}(t,\vec{x})\nonumber\\
&=&w_{1}(t,\vec{x})\exp\{-\frac{i}{\hbar}
[\int_{0}^{t}dt\int d^{3}x w^{\dagger}_{1}(t,\vec{x})
\hat{H}w_{1}(t,\vec{x})\nonumber\\
&&\hspace{1.5 cm}-\int_{0}^{t}dt\int d^{3}x 
w^{\dagger}_{1}(t,\vec{x})i\hbar\partial_{t}w_{1}(t,\vec{x})]\}
\end{eqnarray}
where the last structure is fixed by noting 
$\psi(t,\vec{x})=w_{1}(t,\vec{x})e^{i\phi(t)}$ by 
{\em assumtion}, namely, by the assumption that only
the diagonal component survives for $\psi_{1}(t,\vec{x})$ in 
(70). 

The above formulation makes it clear that $\psi(t,\vec{x})$
is fixed without referring to the equivalence 
class (20) or the notion of the equivalence class of 
Hamiltonians in (21). The geometric term in (74) is determined 
by the hidden local 
symmetry~\footnote{The hidden local symmetry (69) allows us to 
choose a representative 
$\bar{w}_{1}(t,\vec{x})=e^{\alpha(t)}w_{1}(t,\vec{x})$ which 
satisfies the parallel transport condition 
$\int d^{3}\bar{w}^{\dagger}_{1}(t,\vec{x})i\partial_{t}
\bar{w}_{1}(t,\vec{x})=0$. Namely, $\bar{w}_{1}(t,\vec{x})
=\exp[i\int_{0}^{t}dt\int d^{3}x w^{\dagger}_{1}(t,\vec{x})
i\partial_{t}w_{1}(t,\vec{x})]w_{1}(t,\vec{x})$. This 
combination appears in (74), and the quantity manifestly 
invariant under the hidden local symmetry 
\begin{eqnarray}
\bar{w}^{\dagger}_{1}(0,\vec{x})\bar{w}_{1}(T,\vec{x})
=w^{\dagger}_{1}(0,\vec{x})\exp[i\int_{0}^{T}dt\int d^{3}x 
w^{\dagger}_{1}(t,\vec{x})
i\partial_{t}w_{1}(t,\vec{x})]w_{1}(T,\vec{x})
\nonumber
\end{eqnarray}
defines the non-adiabatic phase as holonomy for a cyclic 
evolution of the specific basis vector. Exactly the same 
consideration applies to the adiabatic phase in (17). } 
in (69) with a fixed Hamiltonian 
but without referring to any explicit form of the Hamiltonian. 
The
explicit form of the Hamiltonian is however essential to ensure 
the periodicity of $\psi(t,\vec{x})=\psi_{1}(t,\vec{x})$ up to a
phase for arbitrary $\vec{x}$. The amplitude $\psi(t,\vec{x})$ 
is invariant under the hidden local symmetry 
$w_{1}(t,\vec{x})\rightarrow e^{i\alpha_{1}(t)}w_{1}(t,\vec{x})$
up to a constant phase, $\psi(t,\vec{x})\rightarrow 
e^{i\alpha_{1}(0)}\psi(t,\vec{x})$, and satisfies the linear 
Schr\"{o}dinger equation. 
The quantity 
\begin{eqnarray}
&&\psi^{\dagger}(0,\vec{x})\psi(T,\vec{x})\nonumber\\
&&=w_{1}^{\dagger}(0,\vec{x})w_{1}(T,\vec{x})
\exp\{-\frac{i}{\hbar}
[\int_{0}^{T}dt\int d^{3}x w^{\dagger}_{1}(t,\vec{x})
\hat{H}w_{1}(t,\vec{x})\nonumber\\
&&\hspace{1.5 cm}-\int_{0}^{T}dt\int d^{3}x 
w^{\dagger}_{1}(t,\vec{x})i\hbar\partial_{t}w_{1}(t,\vec{x})]\}
\end{eqnarray}
is thus manifestly invariant under the hidden local symmetry
with a fixed Hamiltonian. Note that the left-hand side of (75) 
is not invariant under 
the equivalence class (20). If one chooses the gauge such that
$w_{1}(0,\vec{x})=w_{1}(T,\vec{x})$ as in our starting 
construction (65), the exponential factor in (75) extracts the 
entire phase from the gauge invariant quantity and , in 
particular, the non-adiabatic phase is given by
\begin{eqnarray}
\beta=\oint dt\int d^{3}x 
w^{\dagger}_{1}(t,\vec{x})i\hbar\partial_{t}w_{1}(t,\vec{x}).
\end{eqnarray}

The hidden local symmetry, which is consistent with the linear 
Schr\"{o}dinger equation, is thus identified as the natural 
origin of the gauge symmetry in the non-adiabatic phase without 
gauge fields. The basis set $\{w_{n}(t,\vec{x})\}$ specify 
the coordinates in the functional space, and they do not satisfy
 the Schr\"{o}dinger equation nor are the eigenvectors of 
$\hat{H}$ in general. The non-adiabatic phase is regarded as a
generalization of the adiabatic phase since it is defined 
without assuming adiabaticity in the sense of the slowness of 
the movement. At the same time, the 
non-adiabatic phase is also regarded as a special case of the 
adiabatic phase in that the exact periodicity of the specific
state $\psi_{1}(t,\vec{x})$ up to a phase is assumed and thus 
the exact adiabaticity in the sense of the absence of 
quantum mixing with other states is assumed. The adiabatic phase
 is rather universal in the sense that one can always define the 
adiabatic phase for any process as long as the (general)
adiabaticity condition is satisfied.

We illustrate this re-formulation of geometric phases in the 
next section.

\section{Explicit examples}
 
\subsection{Adiabatic phase} 
We have already explained that the gauge symmetry in the non -adiabatic phase is not  reduced to that in the adiabatic phase
even in the adiabatic limit. We here analyze the implications of 
this difference. If one takes the equivalence class (20) as a 
gauge symmetry, one is {\em allowed} to choose representatives 
$\bar{\psi}_{1}(t,\vec{x})$ and $\bar{\psi}_{2}(t,\vec{x})$ in
(24), which are gauge invariant up to a constant phase. 
When $\int d^{3}x\psi^{\dagger}(t,\vec{x})i\partial_{t}
\psi(t,\vec{x})$ is identical for  
$\bar{\psi}_{1}(t,\vec{x})$ and $\bar{\psi}_{2}(t,\vec{x})$,
one has 
\begin{eqnarray}
&&c_{1}\bar{\psi}_{1}(t,\vec{x})+
c_{2}\bar{\psi}_{2}(t,\vec{x})\\
&&=e^{i\int_{0}^{t}dt\int d^{3}x\psi_{1}^{\star}(t,\vec{x})
i\partial_{t}\psi_{1}(t,\vec{x})}\{c_{1}\psi_{1}(t,\vec{x})
+c_{2}\psi_{2}(t,\vec{x}) \}\nonumber
\end{eqnarray}
and one can assign the physical meaning to the absolute square
of the superposition. Note that the observable interference 
pattern is {\em unique} and given by 
\begin{eqnarray}
|c_{1}\psi_{1}(t,\vec{x})+c_{2}\psi_{2}(t,\vec{x})|^{2}
\end{eqnarray}
at any moment. The condition (77) needs to be 
satisfied precisely not only at $t=0$ and $t=T$ but also for all
 $t$, since the system is not allowed to go away from quantum 
mechanics in the intermediate stage. See also (45). If one takes
the 
equivalence class (20) literally and chooses representatives 
which satisfy the gauge invariance condition, the interference 
measurement of the non-adiabatic phase thus becomes equivalent 
to the {\em conventional} measurement of interference for a very 
limited set of amplitudes with the constraint (45)
for any $t$, which may also be expressed in terms of an on-shell
value 
$\int d^{3}x\psi^{\dagger}(t,\vec{x})\hat{H}(t)\psi(t,\vec{x})$
with an equivalence class of Hamiltonians. 

 In the adiabatic limit, a superposition of two independent adiabatic solutions
\begin{eqnarray}
&&c_{1}\psi_{1}(\vec{x},t; X(t))+c_{2}\psi_{2}(\vec{x},t; X(t))
\nonumber\\
&&\simeq c_{1}v_{1}(\vec{x};X(t))
\exp\{-\frac{i}{\hbar}\int_{0}^{t}[{\cal E}_{1}(X(t))
-\langle 1|i\hbar\frac{\partial}{\partial t}|1\rangle]dt\}
\nonumber\\
&&+c_{2}v_{2}(\vec{x};X(t))
\exp\{-\frac{i}{\hbar}\int_{0}^{t}[{\cal E}_{2}(X(t))
-\langle 2|i\hbar\frac{\partial}{\partial t}|2\rangle]dt\}
\end{eqnarray}
 satisfies the condition (45) only 
when the  
``dynamical phase'' ${\cal E}_{n}(X(t))$ is identical
\begin{eqnarray}
{\cal E}_{1}(X(t))&=&
\hbar\int d^{3}x\psi_{1}^{\star}(t,\vec{x})
i\partial_{t}\psi_{1}(t,\vec{x})\nonumber\\
&=&\hbar\int d^{3}x\psi_{2}^{\star}
(t,\vec{x})i\partial_{t}\psi_{2}(t,\vec{x})\nonumber\\
&=&{\cal E}_{2}(X(t))
\end{eqnarray}
for the two solutions. This 
gives a sufficient condition to measure the adiabatic 
phase described by $\psi_{n}$, but actually only the 
identical integrated 
\begin{eqnarray}
\int_{0}^{T}dt{\cal E}_{1}(X(t))
=\int_{0}^{T}dt{\cal E}_{2}(X(t))
\end{eqnarray}
is necessary for the direct measurement of the adiabatic phase, 
as is seen in (79). 
The {\em stronger} condition (80) arises from the non-locality 
of $\bar{\psi}$ in $\psi$.  

 An explicit procedure~\cite{berry} to measure the adiabatic 
phase is to 
separate the path of a particle into two in some region of 
space $\vec{x}$; the external parameter $X(t)$ in one of the 
beams, for example, may be chosen to be constant, and the 
superposition of two
 beams is measured later to extract the geometric phase. One 
then controls the "dynamical phase" 
to be identical as in (81) for these two paths. Although the 
stronger condition (80) happens to be
satisfied by an explicit example discussed in~\cite{berry},
only the weaker condition (81) is necessary for the direct 
measurement of the adiabatic phase in interference experiments.
In our re-formulation of the non-adiabatic phase, we encounter only the weaker condition.

\subsection{Spin polarization}

Most of the experimental 
analyses~\cite{bitter, laser,optical, wagh, wagh2} of geometric 
phases are based on the polarization measurements. 
We thus study the model described by  
\begin{eqnarray}
\hat{H}=-\mu\hbar\vec{B}(t)\vec{\sigma}
\end{eqnarray}
where $\vec{\sigma}$ stand for Pauli matrices and $\vec{B}(t)$
is generally a time dependent magnetic field. 

We first briefly comment on the general aspects of the movement 
of  polarization vectors and the holonomy for  spinor basis vectors.    
The most general form of the normalized basis vectors are parameterized as
\begin{eqnarray}
v_{+}(t)=\left(\begin{array}{c}
            \cos\frac{1}{2}\theta(t) e^{-i\varphi(t)}\\
            \sin\frac{1}{2}\theta(t)
            \end{array}\right), \ \ \ \ \ 
v_{-}(t)=\left(\begin{array}{c}
            \sin\frac{1}{2}\theta(t) e^{-i\varphi(t)}\\
            -\cos\frac{1}{2}\theta(t)
            \end{array}\right)
\end{eqnarray}
if one takes the hidden local symmetry (15) into account.
It is also shown that 
\begin{eqnarray}
v^{\dagger}_{+}(t)\vec{\sigma}v_{+}(t)&=&
(\sin\theta\cos\varphi(t), \sin\theta\sin\varphi(t),\cos\theta),
\nonumber\\
&=&-v^{\dagger}_{-}(t)\vec{\sigma}v_{-}(t),
\end{eqnarray}
and
\begin{eqnarray}
\int v_{\pm}^{\dagger}(t)i\partial_{t}v_{\pm}(t)dt&=&-\frac{1}{2}
\int (1\mp\cos\theta)d\varphi+2\pi\nonumber\\
&=&-\frac{1}{2}\Omega_{\pm}
\end{eqnarray}
up to $2n\pi$; $\Omega_{\pm}$ stand for the solid angles drawn 
by the closed movements of unit  polarization vectors in (84).

By using the hidden local symmetry, one may choose a 
representative $\bar{v}_{\pm}(t)=e^{i\alpha(t)}v_{\pm}(t)$
such that 
\begin{eqnarray}
\bar{v}^{\dagger}_{\pm}(t)i\partial_{t}\bar{v}_{\pm}(t)&=&
v_{\pm}^{\dagger}(t)i\partial_{t}v_{\pm}(t)
-\partial_{t}\alpha(t)=0.
\end{eqnarray}
We thus  have
\begin{eqnarray}
\bar{v}_{\pm}(t)=\exp[i\int_{0}^{t}dt 
v(t)_{\pm}^{\dagger}i\partial_{t}v_{\pm}(t)]v_{\pm}(t)
\end{eqnarray}
and 
\begin{eqnarray}
\bar{v}(0)_{\pm}^{\dagger}\bar{v}_{\pm}(T)
=\exp[-i\frac{1}{2}\Omega_{\pm}].
\end{eqnarray}
where $T$ stands for the period of the movement of polarization
vectors.
This shows that the notion of parallel transport (86) and 
holonomy (88), which is analyzed without referring to any 
explicit Hamiltonian, is a notion for the basis 
vectors~\cite{simon, anandan} rather than for the Schr\"{o}dinger 
amplitudes. In the case of Schr\"{o}dinger
amplitudes, one needs to analyze the Schr\"{o}dinger equation 
and the quantum transition between $v_{\pm}(t)$.

We now illustrate our re-formulation of non-adiabatic phases
in the spin polarization phenomena. 

\noindent (i) For the special case~\cite{aharonov}  
\begin{eqnarray}
\vec{B}(t)=(0,0,B)
\end{eqnarray}
in (82) with a constant $B$, one may consider
\begin{eqnarray}
&&\psi_{+}(t)=\cos\frac{1}{2}\theta v_{+}e^{i\hbar\mu Bt/\hbar}
+\sin\frac{1}{2}\theta v_{-}e^{-i\hbar\mu Bt/\hbar},\nonumber\\
&&i\hbar\partial_{t}\psi_{+}(t)=\hat{H}\psi_{+}(t)
\end{eqnarray}
with
\begin{eqnarray}
\hat{H}v_{\pm}=\mp \hbar\mu Bv_{\pm}, \ \ 
v_{+}=\left(\begin{array}{c}
            1\\
            0
            \end{array}\right), \ \ \ \ \ 
v_{-}=\left(\begin{array}{c}
            0\\
            1
            \end{array}\right).
\end{eqnarray}
This model, though quite simple, is conceptually 
important~\cite{aharonov}, and we explain it in some detail.
The amplitude $\psi_{+}(t)$ is written as 
\begin{eqnarray}
\psi_{+}(t)&=&w_{+}(t)\exp[i\hbar\mu Bt/\hbar]\nonumber\\
&=&w_{+}(t)\exp\{-\frac{i}{\hbar}[-\hbar\mu B\cos\theta 
-\hbar\mu B(1-\cos\theta)]t\}
\nonumber\\
&=&w_{+}(t)\exp\{-\frac{i}{\hbar}\int_{0}^{t}dt[
w_{+}(t)^{\dagger}\hat{H}w_{+}(t)
-w_{+}(t)^{\dagger}i\hbar\partial_{t}w_{+}(t)]\}
\end{eqnarray}
with
\begin{eqnarray}
&&w_{+}(t)=\cos\frac{1}{2}\theta v_{+}
+\sin\frac{1}{2}\theta v_{-}e^{-2i\hbar\mu Bt/\hbar},\nonumber\\
&&w_{+}(T)=w_{+}(0),
\nonumber\\
&&w_{+}(t)^{\dagger}i\partial_{t}w_{+}(t)
=\mu B(1-\cos\theta),\nonumber\\
&&w_{+}(t)^{\dagger}\hat{H}w_{+}(t)=-\hbar\mu B\cos\theta
\end{eqnarray}
where $T=\pi/\mu B$.
Similarly, one may define
\begin{eqnarray}
&&\psi_{-}(t)=-\sin\frac{1}{2}\theta v_{+}e^{i\hbar\mu Bt/\hbar}
+\cos\frac{1}{2}\theta v_{-}e^{-i\hbar\mu Bt/\hbar},\nonumber\\
&&i\hbar\partial_{t}\psi_{-}(t)=\hat{H}\psi_{-}(t)
\end{eqnarray}
and 
\begin{eqnarray}
\psi_{-}(t)&=&w_{-}(t)\exp[-i\hbar\mu Bt/\hbar]\nonumber\\
&=&w_{-}(t)\exp\{-\frac{i}{\hbar}[\hbar\mu B\cos\theta
+\hbar\mu B(1-\cos\theta)]t\}
\nonumber\\
&=&w_{-}(t)\exp\{-\frac{i}{\hbar}\int_{0}^{t}dt[
w_{-}(t)^{\dagger}\hat{H}w_{-}(t)
-w_{-}(t)^{\dagger}i\hbar\partial_{t}w_{-}(t)]\}
\end{eqnarray}
with
\begin{eqnarray}
&&w_{-}(t)=-\sin\frac{1}{2}\theta v_{+}e^{2i\hbar\mu Bt/\hbar}
+\cos\frac{1}{2}\theta v_{-},\nonumber\\
&&w_{-}(T)=w_{-}(0),
\nonumber\\
&&w_{-}(t)^{\dagger}i\partial_{t}w_{-}(t)
=-\mu B(1-\cos\theta),\nonumber\\
&&w_{-}(t)^{\dagger}\hat{H}w_{-}(t)=\hbar\mu B\cos\theta.
\end{eqnarray}

We here performed the unitary transformation
\begin{eqnarray}
\left(\begin{array}{c}
      w_{+}(t)\\
      w_{-}(t)
     \end{array}\right)&=&
\left(\begin{array}{cc}
 \cos\frac{1}{2}\theta&\sin\frac{1}{2}\theta e^{-2i\mu Bt}\\
 -\sin\frac{1}{2}\theta e^{2i\mu Bt} &\cos\frac{1}{2}\theta
            \end{array}\right)\left(\begin{array}{c}
            v_{+}\\
            v_{-}
            \end{array}\right)\nonumber\\
&\equiv&U(t)\left(\begin{array}{c}
            v_{+}\\
            v_{-}
            \end{array}\right)
\end{eqnarray}
This means that the expansion of the field variable in second 
quantization is given by 
\begin{eqnarray}
\hat{\psi}(t)&=&\sum_{n}\hat{b}_{n}(t)v_{n}\nonumber\\
&=&\sum_{n}\hat{c}_{n}(t)w_{n}(t)
\end{eqnarray}
with
\begin{eqnarray}
\left(\begin{array}{c}
     \hat{b}_{+}(t)\\
     \hat{b}_{-}(t)
     \end{array}\right)
&\equiv&U^{T}(t)\left(\begin{array}{c}
           \hat{c}_{+}(t)\\
           \hat{c}_{-}(t)
            \end{array}\right)
\end{eqnarray}
where $U^{T}(t)$ stands for the transpose of $U(t)$. 
The effective Hamiltonian $\hat{H}_{eff}(t)$ in (73) in the 
present case is given by
\begin{eqnarray}
\hat{H}_{eff}(t)=\sum_{n=\pm}\hat{c}^{\dagger}_{n}(t)[
w^{\dagger}_{n}(t)\hat{H}w_{n}(t)
-w^{\dagger}_{n}(t)i\hbar\partial_{t}w_{n}(t)]\hat{c}_{n}(t)
\end{eqnarray}
and the off-diagonal terms completely cancel. This formula is
exact and thus non-adiabatic.

The variable $\hat{\psi}(t)$ in (98) is invariant under the 
hidden local gauge symmetry
\begin{eqnarray}
w_{n}(t)\rightarrow e^{i\alpha_{n}(t)}w_{n}(t),\ \ \ \
\hat{c}_{n}(t)\rightarrow e^{-i\alpha_{n}(t)}\hat{c}_{n}(t).
\end{eqnarray}
The expressions of $\psi_{\pm}(t)$ in (92) and (95) are 
invariant under the hidden gauge symmetry up to a constant 
phase. But no symmetry with respect to the equivalence class 
$\{e^{i\alpha_{\pm}(t)}\psi_{\pm}(t)\}$ in (20) nor the 
equivalence class of Hamiltonians (21) appear. We operate on a 
fixed Hamiltonian.

If one recalls that 
\begin{eqnarray}
w^{\dagger}_{+}(t)\vec{\sigma}w_{+}(t)&=&
(\sin\theta\cos\varphi(t), \sin\theta\sin\varphi(t),\cos\theta),
\nonumber\\
&=&-w^{\dagger}_{-}(t)\vec{\sigma}w_{-}(t),
\end{eqnarray}
with $\varphi=-2\mu Bt$, the solid angles $\Omega_{\pm}$ 
subtended by the polarization vectors 
$w^{\dagger}_{\pm}(t)\vec{\sigma}w_{\pm}(t)$ around the z-axis
during a cyclic motion are respectively given by 
\begin{eqnarray}
\int_{0}^{T}dtw_{+}(t)^{\dagger}i\partial_{t}w_{+}(t)
&=&\pi (1-\cos\theta)\nonumber\\
&=&-\frac{1}{2}\int_{0}^{2\pi} d\varphi(1-\cos\theta)
=-\frac{1}{2}\Omega_{+},\nonumber\\
\int_{0}^{T}dtw_{-}(t)^{\dagger}i\partial_{t}w_{-}(t)
&=&-\pi (1-\cos\theta)\nonumber\\
&=&\pi (1+\cos\theta) -2\pi=
-\frac{1}{2}\Omega_{-}
\end{eqnarray}
up to $2n\pi$, where $T=\pi/\mu B$.
 The non-adiabatic phase contained in the 
manifestly gauge invariant $\psi^{\dagger}_{+}(0)\psi_{+}(T)$
, for example, is measured by the interference in
\begin{eqnarray}
|\psi_{+}(T)+\psi_{+}(0)|^{2}
&=&|\psi_{+}(T)|^{2}+|\psi_{+}(0)|^{2}+2{\rm Re}
\psi^{\dagger}_{+}(0)\psi_{+}(T)\nonumber\\
&=&2 + 2\cos[\pi\cos\theta -\frac{1}{2}\Omega_{+}]
\end{eqnarray}
which reproduces the result of Aharonov and 
Anandan~\cite{aharonov}. A separation of the
non-adiabatic phase $-\frac{1}{2}\Omega_{+}$ from the 
``dynamical phase'' 
$\int_{0}^{T}dt w_{+}(t)^{\dagger}\hat{H}w_{+}(t)/\hbar=
-\pi\cos\theta$ is possible if one 
separates the beam into two and later superposes them with a 
suitable Hamiltonian in the second path which cancels the 
``dynamical phase'' $-\pi\cos\theta$ in the first path.
We can thus describe the non-adiabatic phase consistently in 
terms of the hidden local symmetry without referring to the 
equivalence class (20) or the notion of the equivalence class of
Hamiltonians in (21).
\\

\noindent (ii) We next analyze (82) in the case  
\begin{eqnarray}
\vec{B}(t)=B(\sin\theta\cos\varphi(t), \sin\theta\sin\varphi(t),\cos\theta )
\end{eqnarray}
where $\varphi(t)=\omega t$ with constant $\omega$, $B$ and $\theta$.
We then have the effective Hamiltonian in (9) 
\begin{eqnarray}
\hat{H}_{eff}(t)&=&[-\mu\hbar B
-\frac{(1+\cos\theta)}{2}\hbar\omega]\hat{b}^{\dagger}_{+}
\hat{b}_{+}
+[\mu\hbar B-\frac{1-\cos\theta}{2}\hbar\omega]
\hat{b}^{\dagger}_{-}\hat{b}_{-}
\nonumber\\
&-&\frac{\sin\theta}{2}\hbar\omega
[\hat{b}^{\dagger}_{+}\hat{b}_{-}+
\hat{b}^{\dagger}_{-}\hat{b}_{+}]
\end{eqnarray}
with 
\begin{eqnarray}
v_{+}(t)=\left(\begin{array}{c}
            \cos\frac{1}{2}\theta e^{-i\varphi(t)}\\
            \sin\frac{1}{2}\theta
            \end{array}\right), \ \ \ \ \ 
v_{-}(t)=\left(\begin{array}{c}
            \sin\frac{1}{2}\theta e^{-i\varphi(t)}\\
            -\cos\frac{1}{2}\theta
            \end{array}\right)
\end{eqnarray}
which satisfy $\hat{H}(t)v_{\pm}(t)=\mp\mu\hbar Bv_{\pm}(t)$
and the relations
\begin{eqnarray}
v^{\dagger}_{+}(t)i\frac{\partial}{\partial t}v_{+}(t)
&=&\frac{(1+\cos\theta)}{2}\omega
\nonumber\\
v^{\dagger}_{+}(t)i\frac{\partial}{\partial t}v_{-}(t)
&=&\frac{\sin\theta}{2}\omega
=v^{\dagger}_{-}(t)i\frac{\partial}{\partial t}v_{+}(t)
,\nonumber\\
v^{\dagger}_{-}(t)i\frac{\partial}{\partial t}v_{-}(t)
&=&\frac{1-\cos\theta}{2}\omega.
\end{eqnarray}
We next perform a unitary transformation
\begin{eqnarray}
\left(\begin{array}{c}
     \hat{b}_{+}(t)\\
     \hat{b}_{-}(t)
     \end{array}\right)
&=&
\left(\begin{array}{cc}
 \cos\frac{1}{2}\alpha&-\sin\frac{1}{2}\alpha\\
 \sin\frac{1}{2}\alpha &\cos\frac{1}{2}\alpha
            \end{array}\right)
\left(\begin{array}{c}
           \hat{c}_{+}(t)\\
           \hat{c}_{-}(t)
            \end{array}\right)\nonumber\\
&\equiv&U^{T}\left(\begin{array}{c}
           \hat{c}_{+}(t)\\
           \hat{c}_{-}(t)
            \end{array}\right)
\end{eqnarray}
where $U^{T}$ stands for the transpose of $U$. 
The eigenfunctions are transformed to
\begin{eqnarray}
\left(\begin{array}{c}
     w_{+}(t)\\
     w_{-}(t)
     \end{array}\right)
&=&U\left(\begin{array}{c}
           v_{+}(t)\\
           v_{-}(t)
            \end{array}\right)\nonumber\\
&=&
\left(\begin{array}{cc}
 \cos\frac{1}{2}\alpha&\sin\frac{1}{2}\alpha\\
 -\sin\frac{1}{2}\alpha &\cos\frac{1}{2}\alpha
            \end{array}\right)
\left(\begin{array}{c}
           v_{+}(t)\\
           v_{-}(t)
            \end{array}\right)
\end{eqnarray}
or explicitly
\begin{eqnarray}
w_{+}(t)=\left(\begin{array}{c}
            \cos\frac{1}{2}(\theta-\alpha) e^{-i\varphi(t)}\\
            \sin\frac{1}{2}(\theta-\alpha)
            \end{array}\right), \ \ \ \ \ 
w_{-}(t)=\left(\begin{array}{c}
            \sin\frac{1}{2}(\theta-\alpha) e^{-i\varphi(t)}\\
            -\cos\frac{1}{2}(\theta-\alpha)
            \end{array}\right).
\end{eqnarray}
The field variable $\hat{\psi}(t,\vec{x})$ in second quantization
is given by 
\begin{eqnarray}
\hat{\psi}(t,\vec{x})&=&\sum_{n=\pm}\hat{b}_{n}(t)v_{n}(t)
\nonumber\\
&=&\sum_{n=\pm}\hat{c}_{n}(t)w_{n}(t)
\end{eqnarray}
which is invariant under the hidden local symmetry
\begin{eqnarray}
w_{n}(t)\rightarrow e^{i\alpha_{n}(t)}w_{n}(t), \ \ \ \
\hat{c}_{n}(t)\rightarrow e^{-i\alpha_{n}(t)}\hat{c}_{n}(t).
\end{eqnarray}
We also have
\begin{eqnarray}
&&w_{\pm}^{\dagger}(t)\hat{H}w_{\pm}(t)
=\mp \mu\hbar H\cos\alpha\nonumber\\
&&w_{\pm}^{\dagger}(t)i\hbar\partial_{t}w_{\pm}(t)
=\frac{\hbar\omega}{2}(1\pm\cos(\theta-\alpha))
\end{eqnarray}

If one chooses the parameter $\alpha$ in (109) as 
\begin{eqnarray}
\tan\alpha=\frac{\hbar\omega\sin\theta}{2\mu\hbar B+\hbar\omega
\cos\theta}
\end{eqnarray}
one obtains a diagonal effective Hamiltonian
\begin{eqnarray}
\hat{H}_{eff}(t)&=&\sum_{n=\pm}\hat{c}^{\dagger}_{n}
[\mp\mu\hbar B\cos\alpha
-\frac{\hbar\omega}{2}(1\pm\cos(\theta-\alpha))]\hat{c}_{n}
\nonumber\\
&=&\sum_{n=\pm}\hat{c}^{\dagger}_{n}
[w_{n}^{\dagger}(t)\hat{H}w_{n}(t)
-w_{n}^{\dagger}(t)i\hbar\partial_{t}w_{n}(t)]\hat{c}_{n}.
\end{eqnarray}
The above unitary transformation is time-independent and thus 
the effective Hamiltonian is not changed 
$\hat{H}_{eff}(b^{\dagger}_{\pm}(t),b_{\pm}(t))
=\hat{H}_{eff}(c^{\dagger}_{\pm}(t),c_{\pm}(t))$.
We  have the Schr\"{o}dinger amplitudes in (72)
\begin{eqnarray}
\psi_{\pm}(t)
&=&w_{\pm}(t)\exp\{-\frac{i}{\hbar}[\mp\mu\hbar B\cos\alpha
-\frac{\hbar\omega}{2}(1\pm\cos(\theta-\alpha))]t\}
\nonumber\\
&=&w_{\pm}(t)\exp\{-\frac{i}{\hbar}\int_{0}^{t}dt
[w_{\pm}^{\dagger}(t)\hat{H}w_{\pm}(t)
-w_{\pm}^{\dagger}(t)i\hbar\partial_{t}w_{\pm}(t)]\}.
\end{eqnarray}
These expressions are periodic with period  
$T=\frac{2\pi}{\omega}$ up to a phase, and they are exact and 
thus non-adiabatic. From the view point of the diagonalization 
of the Hamiltonian, we have not completely
diagonalized the exact Hamiltonian since $w_{\pm}(t)$ carry
certain time-dependence. These formulas are invariant under 
the hidden local symmetry (113)
up to a constant phase factor, but no invariance under the 
equivalence class (20) nor equivalence
class of Hamiltonians (21) appear. We operate on a fixed 
Hamiltonian.

In the generic case with period $T=\frac{2\pi}{\omega}$, one can
 measure $\psi^{\dagger}_{+}(0)\psi_{+}(T)$, for example, which
 is manifestly invariant under the hidden local symmetry by the 
interference in
\begin{eqnarray}
|\psi_{+}(T)+\psi_{+}(0)|^{2}
&=&2|\psi_{+}(0)|^{2}+2{\rm Re}\psi^{\dagger}_{+}(0)\psi_{+}(T)
\nonumber\\
&=&2+2\cos[(\mu B
\cos\alpha)T
-\frac{1}{2}\Omega_{+}]
\end{eqnarray}
where
\begin{eqnarray}
\Omega_{+}=2\pi [1-\cos(\theta-\alpha)]
\end{eqnarray}
stands for the solid angle drawn by 
$w_{+}^{\dagger}(t)\vec{\sigma}w_{+}(t)$ by noting
\begin{eqnarray}
w_{+}^{\dagger}(t)\vec{\sigma}w_{+}(t)
&=&(\sin(\theta-\alpha)\cos\varphi,\sin(\theta-\alpha)
\sin\varphi,\cos(\theta-\alpha))
\nonumber\\
&=&-w_{-}^{\dagger}(t)\vec{\sigma}w_{-}(t).
\end{eqnarray}
The separation of the non-adiabatic phase and the ``dynamical
phase'' in (118) is achieved by varying the parameters in the 
Hamiltonian, namely, $B$ and $\omega$ in the present case.
The formula (118) however shows that both of the non-adiabatic 
phase and the ``dynamical phase'' depend on these parameters
in a non-trivial way.
In the limit $\hbar\omega\ll \mu\hbar B$, $\alpha\rightarrow 0$
in (115) and the above formula (118) is reduced to the familiar 
adiabatic phase. For $B\rightarrow \rm small$ with fixed 
$T=2\pi/\omega$, $\alpha\rightarrow \theta$ in (115) and the 
geometric phase becomes trivial~\cite{fujikawa2}. More 
generally, in the extreme non-adiabatic limit $\hbar\omega\gg 
\mu\hbar B$, $\alpha\rightarrow \theta$ in (115) and the 
non-adiabatic phase becomes trivial. This fact holds 
independently of an explicit model: The 
phase in (76) becomes trivial $\beta\simeq 0$ in the extreme 
non-adiabatic limit defined by
$\Delta {\cal E} \ll 2\pi\hbar/T$ where $\Delta {\cal E}$ stands
 for the level splitting of a two-level truncation of (9). 
The diagonalization of the dominant geometric term in (9) 
approximately diagonalizes the effective Hamiltonian which gives
 a trivial geometric phase~\cite{fujikawa2}. The 
subtraction of $\int_{0}^{T}dt{\cal E}_{n}(X(t))$ then removes 
the almost degenerate ``dynamical phase'' in (9) and thus 
resulting in the trivial $\beta\simeq 0$.

The eigenfunctions $w_{\pm}(t)$ defined in (111) are 
periodic with period
$T=\frac{2\pi}{\omega}$. By considering the difference of the 
energy factors on the shoulders of the exponential in (117), a 
linear superposition of Schr\"{o}dinger amplitudes 
$\psi_{\pm}(t)$ is periodic with period $T$ up to a phase only 
when 
\begin{eqnarray}
&&T\omega=2\pi n,\nonumber\\
&&T[2\mu B\cos\alpha
+\omega\cos(\theta-\alpha)]=2\pi m
\end{eqnarray}
with two integers $n$ and $m$.
For the generic case, however, a linear superposition of 
$\psi_{\pm}(t)$ is not periodic and  either
 one of $\psi_{\pm}(t)$ is an allowed periodic function with  
period $T=\frac{2\pi}{\omega}$.
For the special case in (121), one may consider a linear 
combination
\begin{eqnarray}
\Psi_{+}(t)&=&\cos\frac{\Theta}{2} \psi_{+}(t)
+\sin\frac{\Theta}{2} \psi_{-}(t)
\nonumber\\
\Psi_{-}(t)&=&-\sin\frac{\Theta}{2} \psi_{+}(t)
+\cos\frac{\Theta}{2} \psi_{-}(t)
\end{eqnarray}
both of which satisfy the Schr\"{o}dinger equation, and one can 
repeat the analysis analogous to (92) and (95) but 
we forgo the details. The movement of the polarization vector in
this case is induced by a superposition of Schr\"{o}dinger 
amplitudes rather than by an attempt to diagonalize the 
evolution operator, and thus it is close to the non-adiabatic 
phase in the original sense of Aharonov and 
Anandan~\cite{aharonov}.

\section{Conclusion}

The notion of rays in the Hilbert space is based on the 
equivalence class
\begin{eqnarray}
\{e^{i\alpha}\psi(t,\vec{x})\}
\end{eqnarray}
with constant phases $\alpha$~\cite{dirac, streater}.
One of the possible generalizations of the above equivalence 
class may be
\begin{eqnarray}
\{e^{i\alpha(t)}\psi(t,\vec{x})\}
\end{eqnarray}
which played a basic role in the definition of the non-adiabatic
phase~\cite{aharonov,anandan2,samuel}. But the origin of this 
gauge symmetry and the consistency
 of imposing the gauge symmetry in the absence of gauge fields 
were not clear.
In particular, a representative, which satisfies the parallel
transport and  gauge invariance conditions, 
\begin{eqnarray}
\bar{\psi}(t,\vec{x})=e^{i\int^{t}_{0} dt\int d^{3}x
\psi^{\dagger}(t,\vec{x})i\partial_{t}\psi(t,\vec{x})}
\psi(t,\vec{x})
\end{eqnarray}
is non-local and non-linear in the Schr\"{o}dinger amplitude
$\psi(t,\vec{x})$, and thus the consistency with the 
superposition principle was not obvious.

We proposed a re-formulation of the non-adiabatic phase on the 
basis
 of the hidden local gauge symmetry~\cite{fujikawa} arising from
 the arbitrariness of the choice of coordinates in the 
functional space. The equivalence class in this case is 
\begin{eqnarray}
\{e^{i\alpha_{n}(t)}w_{n}(t,\vec{x})\}
\end{eqnarray} 
where $\{w_{n}(t,\vec{x})\}$ is a complete orthonormal basis 
set, and this gauge symmetry gives rise to the conventional 
equivalence class (123) for the Schr\"{o}dinger amplitudes. 
The hidden local gauge symmetry  maintains the 
consistency of the non-adiabatic phase with the 
conventional notion of rays and the superposition principle.
This re-formulation clarifies the natural origin of the gauge 
symmetry in both of the adiabatic and non-adiabatic phases 
without gauge fields, and it allows a unified treatment of all 
the geometric phases. 

As for other applications of the second quantized formulation, it has been shown elsewhere that the geometric phase and the quantum anomaly, which have been long considered to be closely
related, in fact have nothing to do with each other~\cite{fujikawa3}.

\bibliography{apssamp}

\end{document}